\documentclass[a4paper,11pt]{article}
\pdfoutput=1 

\usepackage{jcappub} 

\usepackage[T1]{fontenc} 
\usepackage[utf8]{inputenc}
\usepackage{listings}
\usepackage{mathtools}
\usepackage{amsfonts}
\usepackage{amsthm}
\usepackage{amssymb}
\usepackage{eurosym}
\usepackage{amsmath}
\usepackage{epsfig}
\usepackage{graphics}
\usepackage{color}
\usepackage{graphicx}

\title{Teleparallel Equivalent of Lovelock Gravity, Generalizations and Cosmological Applications}

\author[a]{P. A. Gonz\'{a}lez,}
\author[b]{Samuel Reyes}
\author[b]{and Yerko V\'{a}squez}

\affiliation[a]{Facultad de
Ingenier\'{i}a y Ciencias, Universidad Diego Portales, Avenida Ej\'{e}rcito Libertador 441, Casilla 298-V, Santiago, Chile.}
\affiliation[b]{Departamento de F\'isica y Astronom\'ia, Facultad de Ciencias, Universidad de La Serena,\\ Avenida Cisternas 1200, La Serena, Chile.}

\emailAdd{pablo.gonzalez@udp.cl}
\emailAdd{jreyes@userena.cl}
\emailAdd{yvasquez@userena.cl}

\def\be{\begin{equation}}
\def\ee{\end{equation}}
\def\ba{\begin{eqnarray}}
\def\ea{\end{eqnarray}}

\def\bt{\begin{tabular}}
\def\et{\end{tabular}}

\abstract{We consider the teleparallel equivalent of Lovelock gravity and its natural extension, where the action is given by an arbitrary function $f(T_{_{L_1}}, T_{_{L_2}},\cdot \cdot \cdot , T_{_{L_n}})$  of the torsion invariants $T_{_{L_i}}$, which contain higher order torsion terms, and derive its field equations. Then, we consider the special case of $f(T_{_{L_1}}, T_{_{L_2}})$ gravity and study a cosmological scenario by selecting a particular $f(T_{_{L_1}}, T_{_{L_2}})$, and derive the Friedmann equations. Also, we perform a dynamical systems analysis to extract information on the evolution of the cosmological model. Mainly, we find that the model has a very rich phenomenology and can describe the acceleration of the universe at late times.

}

\begin{document}
\maketitle
\flushbottom

\newpage


\section{Introduction}

The teleparallel equivalent of general relativity (TEGR) was proposed by Einstein and is an equivalent formulation of general relativity (GR), where the Weitzenb\"{o}ck connection is used  to define the covariant derivative instead of the Levi-Civita connection, describing a space-time with zero curvature but with non-vanishing torsion which fully encodes the gravitational phenomena \cite{Unzicker:2005in, Hayashi:1979qx}, for an introduction see \cite{Aldrovandi:2013wha}. A natural extension of TEGR is the so-called $f(T)$ gravity, which is represented by a function of the torsion scalar $T$ as Lagrangian density \cite{Ferraro:2006jd,Ferraro:2008ey, Bengochea:2008gz,Linder:2010py}. \\

A genuine advantage of $f(T)$ gravity  
is that the field equations for the vielbein components are second-order differential equations. Recently, it was shown that $f(T)$ gravity has only one extra degree of freedom (d.o.f.), which could be interpreted as a scalar d.o.f. \cite{Ferraro:2018tpu, Ferraro:2018axk}, where the additional scalar field could be related to the proper parallelization of the space-time \cite{Guzman:2019ozl}. Despite this, it was previously found that there is no extra d.o.f. at the level of cosmological perturbations \cite{CP}. It is worth mentioning that some parallelizations have been investigated and the proper vielbein fields responsible for the parallelization process have been found \cite{Fiorini:2013hva}.
Remarkably, it is possible to modify $f(T)$ theory in order to make it manifestly Lorentz invariant \cite{Krssak:2015oua, Golovnev:2017dox}; however, it reduces to $f(T)$ gravity in some local Lorentz frames \cite{Li:2010cg, Weinberg, Arcos:2010gi}. Moreover, $f(T)$ gravity exhibits interesting cosmological implications \cite{Cai:2015emx, Hohmann:2017jao, Nojiri:2017ncd}; for instance, a Born-Infeld $f(T)$ gravity Lagrangian was used to address the physically inadmissible divergences occurring in the standard cosmological Big Bang model, rendering the space-time geodesically complete and powering an inflationary stage without the introduction of an inflaton field \cite{Ferraro:2008ey}. It also is believed that $f(T)$ gravity could be a reliable approach to address the shortcomings of GR at high energy scales \cite{Capozziello:2011et, Nojiri:2017ncd}.  Furthermore, both inflation and the dark energy-dominated stage can be realized in Kaluza-Klein and Randall-Sundrum models, respectively \cite{Bamba:2013fta}. Moreover,  $f(T)$ gravity also admits black hole solutions \cite{solutions}. \\

Other generalizations have been formulated in teleparallel formalism such as scalar-torsion gravity theories \cite{Geng:2011aj, Gonzalez:2014pwa, Kofinas:2015hla, Kofinas:2015zaa, Hohmann:2018rwf, Hohmann:2018vle, Hohmann:2018dqh, Hohmann:2018ijr}, Kaluza-Klein theory for teleparallel gravity  \cite{Geng:2014nfa}, teleparallel equivalent of Gauss-Bonnet gravity \cite{Kofinas:2014owa}, teleparallel equivalent of Lovelock gravity (TELG)  \cite{Gonzalez:2015sha}, teleparallel equivalent of higher dimensional gravity theories \cite{Astudillo-Neira:2017anx} and modified teleparallel gravity models \cite{Bahamonde:2015zma, Bahamonde:2017wwk}. In this work, we will consider a natural extension of 
TELG \cite{Gonzalez:2015sha} by generalizing the Lagrangian of TELG to an arbitrary function of the torsion invariants which contain different powers of the torsion tensor. This is similar to how $f(T)$ gravity modifies the TEGR by considering and arbitrary function of the torsion scalar $T$, which contains quadratic powers of the torsion tensor, and we derive its field equations.
Then, we consider a cosmological scenario in four space-time dimensions described by the spatially flat Friedmann-Lema\^itre-Robertson-Walker (FLRW) metric and a Lagrangian given by $f(T,T_{_{GB}})$, where $T_{_{GB}}$ is a torsion invariant quartic in the torsion tensor (also, it is a topological invariant in four space-time dimensions), and we find the Friedmann equations. The special case $f(T,T_{_{GB}})=T+ \alpha T_{_{GB}}$ corresponds to the teleparallel equivalent of Einstein-Gauss-Bonnet gravity and $\alpha$ is the Gauss-Bonnet coupling constant. Then, by selecting a simple model for $f(T,T_{_{GB}})$ without introducing a new mass scale in the theory,  
we solve the motion equations by performing a dynamical systems analysis to extract information about the evolution of the cosmological model. The dynamical systems approach 
allows us to extract information on the evolution of a cosmological model, regardless of the initial conditions or its specific behavior at intermediate times \cite{Wainwright}.
The Lovelock gravity theory is the most natural extension of GR in higher dimensional space-times that generates second-order field equations  \cite{Lovelock:1971yv}. Remarkably, the action contains terms that appear as corrections to the Einstein-Hilbert action in the context of string theory. Supersymmetric extension, exact black hole solutions, scalar perturbations, thermodynamic, holographic aspects and other properties of Lovelock gravity have been extensively studied over time.\\

It is worth mentioning that in Ref. \cite{Gonzalez:2015sha}, four teleparallel Lagrangians $T_{_{GB}}^{(i)}$ ($i=1,2,3,4$) yielding field equations equivalent to the Gauss-Bonnet gravity and differing among them by boundary terms were constructed. Here, we will consider the simplest of them $T_{_{GB}}^{(1)}$, and 
the remaining Lagrangians can be obtained from $T_{_{GB}}^{(1)}$ by performing integration by parts. Thus, the Lagrangians differ by total derivatives and so yield the same field equations; however, the modified Lagrangians  $f(T,T_{_{GB}}^{(i)})$ with $i=1,2,3,4$ define different gravitational theories and the field equations are not equivalent, except in the special case 
$f(T,T_{_{GB}}^{(i)})=T+\alpha T_{_{GB}}^{(i)}$,
where the theories are also equivalent to the Einstein-Gauss-Bonnet gravity.
The gravity theory considered in this work $f(T,T_{_{GB}}^{(1)})$  is different from the theory analyzed in \cite{Kofinas:2014owa}, which corresponds to $f(T,T_{_{GB}}^{(2)})$. It has been proven that $f(T,T_{_{GB}}^{(2)})$ provides a unified description of the cosmological history from early-times inflation to late-times self-acceleration, without the inclusion of a cosmological constant. Also, the dark energy equation-of-state parameter can be quintessence or phantom-like, or experience the phantom-divide crossing, depending on the parameters of the model \cite{Kofinas:2014daa, Kofinas:2014aka}. We will show that the gravity theory $f(T,T_{_{GB}}^{(1)})$ also has interesting cosmological solutions which are able to describe the acceleration of the universe at late times.\\

The manuscript is organized as follows: In Sec. \ref{FTELG} we give a brief review of TELG. In Sec. \ref{EM}, we modify TELG through its natural extension and the field equations are obtained. Then, in Sec. \ref{FE}, we consider the theory $f(T,T_{_{GB}}^{(1)})$  and we study the cosmological implications by performing a dynamical systems analysis of the Friedmann equations. Finally, we present our conclusions in Sec. \ref{C}.

\section{Fundamentals of the Teleparallel Equivalent of Lovelock Gravity} 
\label{FTELG}

The action of Lovelock gravity written in the language of differential forms reads
\begin{equation}
S= \frac{1}{16 \pi G_D} \int_{\mathcal{M}} \mathcal{L} \, ,
\end{equation}
where $G_D$ is the $D$-dimensional Newton constant and the Lagrangian $D$-form is given by
\begin{equation}
\mathcal{L}=\sum _{n=0}^{[D/2]} \alpha_n \bar{\mathcal{R}}_n \, ,
\end{equation}
where $\alpha_n$ are arbitrary coupling constants, $[D/2]$ means the integer part of $D/2$ and each of the terms $\bar{\mathcal{R}}_n$ corresponds to the dimensional extension of the Euler density in $2n$ dimensions and is given by
\begin{eqnarray} \label{lovelock}
\notag \bar{\mathcal{R}}_n &=& \frac{1}{(D-2n)!} \epsilon_{a_1 a_2 \cdot \cdot \cdot a_D} \bar{R}^{a_1 a_2} \wedge \bar{R} ^{a_3 a_4} \wedge \cdot \cdot \cdot \wedge \bar{R}^{a_{2n-1}a_{2n}} \wedge e^{a_{2n+1}} \wedge \cdot \cdot \cdot \wedge e^{a_D} \\
&=& \frac{1}{(D-2n)!}\epsilon_{a_1 a_2 \cdot\cdot\cdot  a_D} \Bigg[ \prod_{i=1}^{n} \bar{R} ^{a_{2i-1} a_{2i}}\Bigg] \wedge \Bigg[\prod_{i=2n+1}^{D} e^{a_i} \Bigg]\,,
\end{eqnarray}
where the symbol $\prod$ denotes the wedge product of a sequence of differential forms:
\begin{equation}
\prod_{i=m}^{n} \theta_i = \theta_m \wedge \theta_{m+1} \wedge \cdot \cdot \cdot \wedge \theta_{n} \,\,\, \text{for} \,\,\, n>m\,, \prod_{i=m}^{n} \theta_i = \theta_m \,\,\,\, \text{for} \,\,\, n=m \,\,\, \text{and} \,\,\,   \prod_{i=m}^n \theta_i=1 \,\,\, \text{for} \,\,\, n<m \,,
\end{equation}
for some $p$-forms $\theta_i$.
The indices $a_i$ can take the values $a_i=0,1, \cdot \cdot \cdot, D-1$, and we are using the convention $\epsilon_{01\cdot \cdot \cdot D-1}=1$. Only the terms with $n< D/2$ contribute to the field equations, and the terms with $n=D/2$ are topological invariants in $2n$ dimensions. $\bar{\mathcal{R}}_0$ is the cosmological term when $\alpha_0 \neq 0$. $\bar{R}^a_{\,\ b}=d \bar{\omega}^a_{\,\ b}+\bar{\omega}^a_{\,\ c}\wedge \bar{\omega}^c_{\,\ b}$ is the curvature 2-form corresponding to the torsionless Levi-Civita spin connection $\bar{\omega}^{ab}$, while the curvature 2-form of an arbitrary spin connection $\omega^{ab}$ is defined by
\begin{equation}
R^a_{\,\ b}=d\omega^a_{\,\ b}+\omega^a_{\,\ c}\wedge\omega^c_{\,\ b}~,
\end{equation}
and the torsion 2-form is defined by
\begin{equation}
T^a=de^a+\omega^a_{\,\ b}\wedge e^b~.
\end{equation}
The vielbein field $e^{a}$ forms an orthonormal basis for the cotangent space at each point of the manifold, while the vector field $\mathbf{e}_{a}$ forms an orthonormal basis in the tangent space, being dual to $e^{a}$, $e^{a}(\mathbf{e}_b)=\delta ^a _b$, where $\delta^{a}_b$ is the Kronecker delta. We use a bold letter for the vector fields $\mathbf{e}_a$, while $e_a=\eta_{ab} e^b$ are 1-forms.
In a gravity theory with non-null curvature and non-null torsion, the vielbein $e^a$ and the spin connection $\omega^a_{\,\ b}$ fields are dynamical variables, see for instance \cite{Espiro:2014uda} where cosmological models with propagating torsion were considered. On the other hand, for null torsion, the spin connection corresponds to the Levi-Civita spin connection $\bar{\omega}^{ab}$, and it only depends on $e^a$.
The arbitrary spin connection can be split as
\begin{equation}\label{contortion}
\omega^{ab}= \bar{\omega}^{ab}+K^{ab} \, ,
\end{equation}
where $K^{ab}$ is the contortion 1-form, which is related to the torsion 2-form by $T^a = K^{a} _{\,\,\, b} \wedge e^{b}$. On the other hand, the exterior covariant derivative $D$ of the connection $\omega^{ab}$ acts on a set of $p$-forms $\phi^a _{\,\ b}$ as $D\phi^a_{\,\ b}=d\phi^a_{\,\ b}+\omega^a_{\,\ c}\wedge\phi^c_{\,\ b}-(-1)^p\phi^a_{\,\ c}\wedge\omega^c_{\,\ b}$, while that the covariant derivative $\bar{D}$ is defined in a similar manner for the Levi-Civita connection $\bar{\omega}^{ab}$, being $\bar{D}e^a=de^a+\bar{\omega} ^a_{\,\,\, b} \wedge e^b=0$. Using (\ref{contortion}) it is possible to write the curvature in terms of the Riemannian curvature ($\bar{R}^{ab}$) and the contortion as
\begin{equation} \label{curvatura}
R^a_{\,\,\, b}=\bar{R}^a_{\,\,\, b}+\bar{D}K^a_{\,\,\, b}+K^a_{\,\ c}\wedge K^c_{\,\,\,b}~,
\end{equation}
where the covariant derivative $\bar{D}$ acts on $K^a_{\,\,\, b}$ by $\bar{D}K^a_{\,\ b}=dK^a_{\,\ b}+\bar{\omega}^a_{\,\ c}\wedge K^c_{\,\ b}+K^a_{\,\ c}\wedge \bar{\omega}^c_{\,\ b}$. In a similar form $D$ acts on $K^a_{\,\,\,b}$ by $DK^a_{\,\ b}=dK^a_{\,\ b}+\omega^a_{\,\ c}\wedge K^c_{\,\ b}+K^a_{\,\ c}\wedge\omega^c_{\,\ b}$.


The teleparallel equivalent of the generic $\bar{\mathcal{R}}_n$ term (\ref{lovelock}) was constructed in \cite{Gonzalez:2015sha}, and it is given by
\begin{eqnarray} \label{lagrangian}
\notag \tau_{_{L_n}} &=& \frac{1}{(D-2n)!}\epsilon_{a_1 a_2 \cdot\cdot\cdot  a_D}(K^{a_1 c_1}\wedge K_{c_1}^{\,\,\,a_2}-DK^{a_1 a_2})\wedge\cdot\cdot\cdot \, \wedge (K^{a_{2n-3} c_{2n-3}}\wedge K_{c_{2n-3}}^{\,\,\,\,\,\,\,\,a_{2n-2}}-DK^{a_{2n-3} a_{2n-2}}) \\
\notag && \wedge ((1-2n)K^{a_{2n-1} c_{2n-1}}\wedge K^{\,\,\,\,\,\,\,\, a_{2n}}_{ c_{2n-1}}-(1-n)DK^{a_{2n-1} a_{2n}})\wedge e^{a_{2n+1}}\wedge\cdot\cdot\cdot\, \wedge e^{a_D} \\
\notag &=& \frac{1}{(D-2n)!}\epsilon_{a_1 a_2 \cdot\cdot\cdot  a_D} \Bigg[ \prod_{i=1}^{n-1} (K^{a_{2i-1} c_{2i-1}} \wedge K_{c_{2i-1}}^{\,\,\,\,\,\,\,\, a_{2i}}-DK^{a_{2i-1} a_{2i}} ) \Bigg] \wedge \Big((1-2n)K^{a_{2n-1} c_{2n-1}}\wedge K^{\,\,\,\,\,\,\,\, a_{2n}}_{ c_{2n-1}}- \\
&& (1-n)DK^{a_{2n-1} a_{2n}}\Big)\wedge \Bigg[\prod_{i=2n+1}^{D} e^{a_i} \Bigg] \, ,
\end{eqnarray}
which is based only on the space-time torsion, and differs from $\bar{\mathcal{R}}_n$ only by a total derivative (boundary term):
\begin{equation} \label{bou}
\bar{\mathcal{R}}_n = \tau_{_{L_n}}-\frac{1}{(D-2n)!}d \left( n \epsilon_{a_1 a_2 \cdot \cdot \cdot a_D} K^{a_1 a_2} \wedge \Bigg[ \prod_{i=1}^{n-1}\bar{R} ^{a_{2i+1} a_{2i+2}} \Bigg] \wedge \Bigg[\prod_{i=2n+1}^{D} e^{a_i} \Bigg] \right) \, .
\end{equation}
Therefore, the same field equations are obtained from both Lagrangians and, such as $\bar{\mathcal{R}}_n$, $\tau_{_{L_n}}$ also is a topological invariant in $2n$ space-time dimensions; thus, the theory defined by
\begin{equation} \label{action1}
S= \frac{1}{16 \pi G_D} \int_{\mathcal{M}}  \mathcal{L}_{_{TELG}}\, ,
\end{equation}
with $\mathcal{L}_{_{TELG}}=\sum_{n=0}^{[D/2]} \alpha_n \tau_{_{L_n}}$  is called TELG.
Torsion invariants of degree $2n$ in the torsion tensor can be constructed as $T_{_{L_n}}=- \star \tau_{_{L_n}}$, where $\star$ denotes the Hodge star operator. These invariants will be used in the next section to construct a generalization of TELG, where the Lagrangian is given by an arbitrary function of these invariants $f(T_{_{L_1}}, T_{_{L_2}},\cdot \cdot \cdot , T_{_{L_n}})$. From (\ref{bou}) we obtain $\tau_{_{L_0}}=\bar{\mathcal{R}}_0$ is the cosmological term and $T_{_{L_0}}=1$. We have not included $T_{_{L_0}}$ as an argument for the arbitrary function $f$ due to it being just a constant term.

The Lagrangian of TEGR (in $D$ space-time dimensions) can be obtained from (\ref{lagrangian}) with $n=1$ and is given by
\begin{eqnarray} \label{scalar}
\notag \tau_{_{L_1}} &=& - \frac{1}{(D-2)! } \epsilon _{a_1 a_2 ... a_D} K^{a_1 c_1} \wedge K_{c_1} ^{\,\,\,a_2} \wedge e^{a_3} \wedge \cdot \cdot \cdot  \wedge e^{a_D} \\
&& = T_{_{GR}} \, e^0 \wedge e^1 \wedge \cdot \cdot \cdot \wedge e^{D-1} \, ,
\end{eqnarray}
where we have defined $T_{_{GR}} \equiv T_{_{L_1}} = -\star \tau_{_{L_1}}$, while the Lagrangian of teleparallel equivalent of Gauss-Bonnet gravity corresponds to $n= 2$ in (\ref{lagrangian}) and it reads
 \begin{eqnarray} \label{gb}
\notag  \tau_{_{L_2}} &=& \frac{1}{(D-4)!} \epsilon_{a_1 a_2 \cdot\cdot\cdot  a_D}(K^{a_1 c_1}\wedge K_{c_1}^{\,\,\,a_2}-DK^{a_1 a_2})\wedge (-3K^{a_3 c_3}\wedge K^{\,\,\,\,\,\,\,\, a_4}_{ c_3}+DK^{a_3 a_4})\wedge  \\
\notag   && e^{a_5}\wedge\cdot\cdot\cdot\, \wedge e^{a_D} \\
  &=& T_{_{GB}}^{(1)} \, e^0 \wedge e^1 \wedge \cdot \cdot \cdot \wedge e^{D-1} \, ,
\end{eqnarray}
where we have defined $T_{_{GB}}^{(1)} \equiv T_{_{L_2}} = -\star \tau_{_{L_2}}$, and so on for the other terms $n=3, 4, \cdot \cdot \cdot$.
From (\ref{bou}) we find that $\tau_{_{L_1}}$ and $\tau_{_{L_2}}$ deviate from the Einstein-Hilbert term and the Gauss-Bonnet term, respectively, by the total derivatives:
\begin{eqnarray}
\bar{\mathcal{R}}_1 &=& \tau_{_{L_1}}-\frac{1}{(D-2)!}d \left(\epsilon_{a_1 a_2 \cdot \cdot \cdot a_D} K^{a_1 a_2} \wedge e^{a_3} \wedge e^{a_4} \cdot \cdot \cdot \wedge e^{a_D}  \right) \\
\bar{\mathcal{R}}_2 &=& \tau_{_{L_2}} - \frac{1}{(D-4)!} d \left( 2 \epsilon_{a_1 a_2 \cdot \cdot \cdot a_D} K^{a_1 a_2} \wedge \bar{R} ^{a_3 a_4} \wedge e^{a_5} \wedge \cdot \cdot \cdot \wedge e^{a_D} \right) \,.
\end{eqnarray}
It is worth mentioning that by performing integration by parts to the Lagrangian $\tau_{_{L_2}}$, other Lagrangians 
are obtained and these are given by \cite{Gonzalez:2015sha}
\begin{eqnarray}
 \notag   \Theta^{(2)}&=&\frac{1}{(D-4)!}\epsilon_{a_1 a_2 \cdot \cdot \cdot a_D}(-2K^{a_1 a_2} \wedge K^{a_3}_{\,\,\,\,\, c} \wedge K^{c}_{\,\,\,\,\, d} \wedge K^{d a_4}+2K^{a_1 a_2}\wedge D K^{a_3}_{\,\,\,\,\, c} \wedge K^{c a_4} + \\
\notag && K^{a_1}_{\,\,\,\,\, c} \wedge K^{c a_2} \wedge K^{a_3}_{\,\,\,\,\, d} \wedge K^{d a_4}) \wedge e^{a_5} \wedge \cdot \cdot \cdot \wedge e^{a_D} \\
&=& T_{_{GB}}^{(2)} \, e^0 \wedge e^1 \wedge \cdot \cdot \cdot \wedge e^{D-1}  
\,, \\
 \notag \Theta^{(3)}&=& \frac{1}{(D-4)!}\epsilon_{a_1 a_2 \cdot \cdot \cdot a_D}(-D K^{a_1 a_2} \wedge DK^{a_3 a_4}+6 DK^{a_1 a_2} \wedge K^{a_3}_{\,\,\,\,\, c} \wedge K^{c a_4} \\
 \notag  && -7K^{a_1}_{\,\,\,\,\, c} \wedge K^{c a_2} \wedge K^{a_3}_{\,\,\,\,\, d} \wedge K^{d a_4}- 4 K^{a_1 a_2} \wedge D K^{a_3}_{\,\,\,\,\, d} \wedge K^{d a_4}+ 8 K^{a_1 a_2} \wedge K^{a_3}_{\,\,\,\,\, c} \wedge \\
\notag  && K^{c}_{\,\,\,\,\, d} \wedge K^{d a_4}) \wedge e^{a_5} \wedge \cdot \cdot \cdot \wedge e^{a_D} \\
  &=& T_{_{GB}}^{(3)} \, e^0 \wedge e^1 \wedge \cdot \cdot \cdot \wedge e^{D-1}  \, , \\
\notag  \Theta^{(4)} &=&  \frac{1}{(D-4)!}\epsilon_{a_1 a_2 \cdot \cdot \cdot a_D}( 6  K^{a_1 a_2} \wedge K^{a_3}_{\,\,\,\,\, c} \wedge K^{c}_{\,\,\,\,\, d} \wedge K^{d a_4} -2 K^{a_1 a_2} \wedge D K^{a_3}_{\,\,\,\,\, d} \wedge K^{d a_4} -  \\
\notag && 3 K^{a_1}_{\,\,\,\,\, c} \wedge K^{c a_2} \wedge K^{a_3}_{\,\,\,\,\, d} \wedge K^{d a_4}+2 D K^{a_1 a_2} \wedge K^{a_3}_{\,\,\,\,\, c} \wedge K^{c a_4} ) \wedge e^{a_5} \wedge \cdot \cdot \cdot \wedge e^{a_D} \\
 &=& T_{_{GB}}^{(4)} \,  e^0 \wedge e^1 \wedge \cdot \cdot \cdot \wedge e^{D-1} \,  .
\end{eqnarray}
where $T_{_{GB}}^{(2)}=- \star \Theta^{(2)}$, $T_{_{GB}}^{(3)}= -\star \Theta^{(3)}$ and $T_{_{GB}}^{(4)}=-\star \Theta^{(4)}$. In \cite{Gonzalez:2015sha} there is a mistake in the sign of the fourth term of $\Theta ^{(3)}$, we have fixed it here. These Lagrangians, such as $\tau_{_{L_2}}$, also yield the same field equations of Gauss-Bonnet gravity, because each of them differs from $\bar{\mathcal{R}}_2$ by the following total derivatives
\begin{eqnarray}
\notag \bar{\mathcal{R}}_2 &=& \Theta^{(2)} - \frac{1}{(D-4)!} d \big( 2 \epsilon_{a_1 a_2 \cdot \cdot \cdot a_D} K^{a_1 a_2} \wedge \bar{R} ^{a_3 a_4} \wedge e^{a_5} \wedge \cdot \cdot \cdot \wedge e^{a_D} + \\
\notag && \epsilon_{a_1 a_2 \cdot \cdot \cdot a_D} K^{a_1 a_2} \wedge \bar{D} K^{a_3 a_4} \wedge e^{a_5} \wedge \cdot \cdot \cdot  \wedge e^{a_D}\big) \\
\notag &=& \Theta^{(3)} - \frac{1}{(D-4)!}d \big(2  \epsilon_{a_1 a_2 \cdot \cdot \cdot a_D} K^{a_1 a_2} \wedge \bar{R} ^{a_3 a_4} \wedge e^{a_5} \wedge \cdot \cdot \cdot \wedge e^{a_D} + \\
\notag && 2 \epsilon_{a_1 a_2 \cdot \cdot \cdot a_D} K^{a_1 a_2} \wedge K ^{a_3 c} \wedge K_{c}^{\,\,\, a_4} \wedge e^{a_5} \wedge \cdot \cdot \cdot \wedge e^{a_D}  \big) \\
\notag &=& \Theta^{(4)} - \frac{1}{(D-4)!} d \big( 2 \epsilon_{a_1 a_2 \cdot \cdot \cdot a_D} K^{a_1 a_2} \wedge \bar{R} ^{a_3 a_4} \wedge e^{a_5} \wedge \cdot \cdot \cdot \wedge e^{a_D} + \\
\notag && \epsilon_{a_1 a_2 \cdot \cdot \cdot a_D} K^{a_1 a_2} \wedge \bar{D} K^{a_3 a_4} \wedge e^{a_5} \wedge \cdot \cdot \cdot  \wedge e^{a_D} + \\
&& 2 \epsilon_{a_1 a_2 \cdot \cdot \cdot a_D} K^{a_1 a_2} \wedge K ^{a_3 c} \wedge K_{c}^{\,\,\, a_4} \wedge e^{a_5} \wedge \cdot \cdot \cdot \wedge e^{a_D}
\big) \,.
\end{eqnarray}

In the following, we will impose the Weitzenb{\"o}ck connection by setting $\omega^{a b}=0$ in order to simplify further the above expressions, in this case the exterior covariant derivative $DK^{a b}$ reduces to the exterior derivative $dK^{a b}$. As mentioned,
the Lagrangian of teleparallel equivalent of Gauss-Bonnet gravity constructed in \cite{Kofinas:2014owa} corresponds to $T_{_{GB}}^{(2)}$. The modified Lagrangians $f(T_{_{GR}},T_{_{GB}}^{(i)})$ with $i=1,2,3,4$ define different gravitational theories, and they are equivalent only when $f(T_{_{GR}},T_{_{GB}}^{(i)})=T_{_{GR}}+  \alpha T_{_{GB}}^{(i)}$, which corresponds to Einstein-Gauss-Bonnet gravity. In the next section we will study the cosmological implications of the theory given by $f(T_{_{GR}},T_{_{GB}}^{(1)})$, where $T_{_{GR}}$ and $T_{_{GB}}^{(1)}$ are given by (\ref{scalar}) and (\ref{gb}), respectively.

\section{Generalizations of the Teleparallel Equivalent of Lovelock Gravity} 
\label{EM}

A natural generalization of TELG is given by the action
\begin{equation}
\label{MTELG}
  S=\frac{1}{16 \pi G_D} \int_{\mathcal{M}} d^D x e f(T_{_{L_1}}, T_{_{L_2}}, \cdot \cdot \cdot , T_{_{L_n}})+ S_m \,,
\end{equation}
where $e=det(e^a_{\,\,\, \mu})= \sqrt{-g}$, the terms $T_{_{L_n}}=- \star \tau_{_{L_n}}$ are torsion invariants of degree $2n$ in the torsion tensor, which for $n \geq \left[ \frac{D+1}{2} \right]$ are also topological invariants in $D$ space-time dimensions, and $S_m$ is an action for matter. TELG corresponds to the special case $f(T_{_{L_1}}, T_{_{L_2}}, \cdot \cdot \cdot , T_{_{L_n}})= \alpha_0+ \sum_{m=1}^n \alpha_m T_{_{L_m}}$, where $\alpha_0$ corresponds to the cosmological constant and $\alpha_m$ are arbitrary coupling constants. The equations of motion can be 
obtained by varying the action (\ref{MTELG}) with respect to the dynamical fields $e^a$, which yields
\begin{equation}
\label{action}
    16 \pi G_D \delta S = \int_{\mathcal{M}} d^Dx \left( f \delta e + e \sum_{i=1}^n \frac{\partial f }{\partial T_{_{L_i}}} \delta T_{_{L_i}} \right) +16 \pi G_D \delta S_m \,.
\end{equation}
However, by using differential forms, (\ref{action}) can be written as
\begin{equation}
    16 \pi G_D \delta S = \int_{\mathcal{M}} \sum_{i=1}^n f_{T_{_{L_i}}} \delta \tau_{_{L_i}} + \int_{\mathcal{M}} d^Dx \delta e \left( f- \sum_{i=1}^n T_{_{L_i}} f_{T_{_{L_i}}} \right)+16 \pi G_D\int_{\mathcal{M}} \delta{\mathcal{L}_m} \,  ,
\end{equation}
where we have defined $f_{T_{_{L_i}}} \equiv \frac{\partial f }{\partial T_{_{L_i}}}$, and 
$\mathcal{L}_m$ is the matter Lagrangian $D$-form. The term inside the first integral is found to be
\begin{equation}
f_{T_{_{L_n}}} \delta \tau_{_{L_n}}= \delta K_{ab} \wedge H^{ab (n)}+ \delta e^a \wedge h_a^{(n)} \,,
\end{equation}
where
\begin{eqnarray}
\notag H^{ab(n)} &=& \frac{2 f_{T_{_{L_n}}}}{(D-2n)!} \Bigg(    (1-2n)\epsilon ^a_{\,\,\, a_1 a_2 \cdot \cdot \cdot  a_{D-1}} \Bigg[\prod_{i=1}^{n-1} \left(K^{a_{2i-1} c_{2i-1}} \wedge K_{c_{2i-1}}^{\,\,\,\,\,\,\, a_{2i}} -dK^{a_{2i-1} a_{2i}} \right) \Bigg] \wedge K^{b a_{2n-1}} +  \\
\notag &&  (n-1)\epsilon^{a}_{\,\,\, a_1 a_2 \cdot \cdot \cdot a_{D-1}}  \Bigg[\prod_{i=1}^{n-2} \left( K^{a_{2i-1} c_{2i-1}} \wedge K_{c_{2i-1}}^{\,\,\,\,\,\,\, a_{2i}} -dK^{a_{2i-1} a_{2i}}\right) \Bigg] \wedge \Big((1-2n)K^{a_{2n-3} c_{2n-3}} \wedge K_{c_{2n-3}}^{\,\,\,\,\,\,\, a_{2n-2}} \\
\notag && -(1-n) dK^{a_{2n-3} a_{2n-2}} \Big) \wedge K^{b a_{2n-1}} \Bigg) \wedge \prod_{i=2n}^{D-1} e^{a_i} \\
\notag && - d \Bigg( \frac{f_{T_{_{L_n}}}}{(D-2n)!} \Bigg[ (1-n) \epsilon^{ab}_{\,\,\,\,\, a_1 a_2 \cdot \cdot \cdot a_{D-2}}  \Bigg[ \prod_{i=1}^{n-1}  \left( K^{a_{2i-1} c_{2i-1}} \wedge K_{c_{2i-1}}^{\,\,\,\,\,\,\, a_{2i}} -dK^{a_{2i-1} a_{2i}} \right) \Bigg] + \\
\notag && (n-1)\epsilon^{ab}_{\,\,\,\,\, a_1 a_2 \cdot \cdot \cdot a_{D-2}} \Bigg[ \prod_{i=1}^{n-2} \left(  K^{a_{2i-1} c_{2i-1}} \wedge K_{c_{2i-1}}^{\,\,\,\,\,\,\, a_{2i}}-dK^{a_{2i-1} a_{2i}} \right) \Bigg] \wedge \Big( (1-2n)K^{a_{2n-3} c_{2n-3}} \wedge K_{c_{2n-3}}^{\,\,\,\,\,\,\, a_{2n-2}} \\
&& - (1-n) dK^{a_{2n-3} a_{2n-2}}  \Big)   \Bigg]  \wedge \prod_{i=2n-1}^{D-2} e^{a_i} \Bigg)
\end{eqnarray}
and
\begin{eqnarray}
\notag h_a^{(n)} &=&\frac{(D-2n)}{(D-2n)!} f_{T_{_{L_n}}}\epsilon_{a a_1 a_2 \cdot \cdot \cdot a_{D-1}} \Bigg[ \prod_{i=1}^{n-1} \left(K^{a_{2i-1} c_{2i-1}} \wedge K_{c_{2i-1}}^{\,\,\,\,\,\,\, a_{2i}} -  dK^{a_{2i-1} a_{2i}} \right) \Bigg] \wedge \\
&& \left( (1-2n)K^{a_{2n-1} c_{2n-1}} \wedge K_{c_{2n-1}}^{\,\,\,\,\,\,\, a_{2n}}-(1-n) dK^{a_{2n-1} a_{2n}} \right)\wedge \prod_{i=2n+1}^{D-1} e^{a_i} \,.
\end{eqnarray}
The first two terms $n=1,2$ of the above expressions are given respectively by
\begin{eqnarray}
\notag H^{ab(1)} &=&-\frac{2 f_{T_{_{L_1}}}}{(D-2)!}\epsilon^a_{\,\,\, a_1 a_2 \cdot \cdot \cdot a_{D-1}} K^{b a_1} \wedge e^{a_2} \wedge \cdot \cdot \cdot \wedge e^{a_{D-1}}\, , \\
\notag H^{ab(2)} &=& \frac{2 f_{T_{_{L_2}}}}{(D-4)!} \Big(-3 \epsilon^a_{\,\,\, a_1 a_2 \cdot \cdot \cdot a_{D-1}} (K^{a_1 c_1} \wedge K_{c_1}^{\,\,\,\,\, a_2}-dK^{a_1 a_2})\wedge K^{b a_3} + \\
\notag &&  \epsilon^{a}_{\,\,\, a_1  a_2 \cdot \cdot \cdot  a_{D-1}} (-3 K^{a_1 c_1} \wedge K_{c_1}^{\,\,\,\,\, a_2}+dK^{a_1 a_2})\wedge K^{b a_3 }  \Big) \wedge e^{a_4} \wedge \cdot \cdot \cdot \wedge e^{a_{D-1}} - \\
\notag && d \Big( \frac{f_{T_{_{L_2}}}}{(D-4)!}\Big[ -\epsilon^{ab}_{\,\,\,\,\, a_1 a_2 \cdot \cdot \cdot a_{D-2}} (K^{a_1 c_1} \wedge K_{c_1}^{\,\,\,\,\, a_2}-dK^{a_1 a_2})  + \\
\notag && \epsilon^{ab}_{\,\,\,\,\, a_1 a_2 \cdot \cdot \cdot a_{D-2}} (-3 K^{a_1 c_1} \wedge K_{c_1}^{\,\,\,\,\, a_2}+ dK^{a_1 a_2})  \Big] \wedge e^{a_3} \wedge \cdot \cdot \cdot \wedge e^{a_{D-2}} \Big)
\end{eqnarray}
and 
\begin{eqnarray}
\notag h_a^{(1)} &=& -\frac{D-2}{(D-2)!}  f_{T_{_{L_1}}} \epsilon_{a a_1 \cdot \cdot \cdot a_{D-1}} K^{a_1 c_1} \wedge K_{c_1}^{\,\,\,\,\, a_2} \wedge e^{a_3} \wedge \cdot \cdot \cdot e^{a_{D-1}} \,    ,\\
\notag h_a^{(2)} &=& \frac{D-4}{(D-4)!} f_{T_{_{L_2}}} \epsilon_{a a_1 \cdot \cdot \cdot a_{D-1}} (K^{a_1 c_1} \wedge K_{c_1}^{\,\,\,\,\, a_2}-dK^{a_1 a_2})\wedge (-3 K^{a_3 c_3} \wedge K_{c_3}^{\,\,\,\,\, a_4}+ dK^{a_3 a_4}) \wedge e^{a_5} \wedge \cdot \cdot \cdot \wedge e^{a_{D-1}}\, .
\end{eqnarray}

On the other hand, the variation of the matter Lagrangian yields $\delta \mathcal{L}_{m}= \delta e^{a} \frac{\delta \mathcal{L}_m}{\delta e^{a}}= \delta e^a T_{ab} \star e^b$, where $ T_{ab} \star e^b$ is the energy-momentum $D-1$-form.
\newline

The field equations obtained by varying the action (\ref{MTELG}) are
\begin{eqnarray}
\notag && \sum_{i=1}^n \Bigg( \mathcal{L}_{\mathbf{e}_b}H^{[ab(i)]} - i_{\mathbf{e}_b} \mathcal{L}_{\mathbf{e}_c} (e^c \wedge H^{[ab(i)]}+e^a \wedge H^{[cb(i)]}) -\frac{1}{2} C^d_{\,\,\, cb} i_{\mathbf{e}_d} (e^a \wedge H^{cb(i)})+2C_{(dc)}^{\,\,\,\,\,\,\,\, a} i_{\mathbf{e}_b} (e^c \wedge H^{[db(i)]})   \\
&& +\frac{1}{2} T^a_{\,\,\, bc} H^{bc(i)} +h^{a (i)}\Bigg)  +(f-\sum_{i=1}^n T_{_{L_i}} f_{T_{_{L_i}}}) \theta^a =-16 \pi G_D T^{a}_{\,\,\, b} \star e^{b} \, ,
\end{eqnarray}
where $\mathcal{L}_{\mathbf{e}_a}$ denotes the Lie derivative with respect to the vector field $\mathbf{e}_a$, $H^{[ab(i)]}= \frac{1}{2} (H^{ab(i)}-H^{ba(i)})$, $C_{(dc)}^{\,\,\,\,\,\,\,\, a}=\frac{1}{2} (C_{dc}^{\,\,\,\,\,\,\,\, a}+C_{cd}^{\,\,\,\,\,\,\,\, a})$, $i_{\mathbf{e}_a}$ is the interior product and satisfies properties such as $i_{\mathbf{e}_a}(e^b)=\delta^b_a$ and $i_{\mathbf{e}_a} (e^b \wedge e^c)= i_{\mathbf{e}_a}(e^b) e^c- i_{\mathbf{e}_a}(e^c) e^b$, $C^a_{\,\,\, bc}$ are the structure coefficients or coefficients of anholonomy, defined from the commutation relation $[\mathbf{e}_b,\mathbf{e}_c]=C^a_{\,\,\, bc} \mathbf{e}_a $, $T^{a}_{\,\,\, bc}$ are the components of the torsion tensor when expressed in the orthonormal base $T^{a}=\frac{1}{2} T^{a}_{\,\,\, bc} e^{b}\wedge e^{c}$ and $\theta_{a}= i_{\mathbf{e}_a} (e^0 \wedge \cdot \cdot \cdot \wedge e^{D-1})$.

An interesting feature of the field equations 
of $f(T_{_{L_1}}, T_{_{L_2}}, \cdot \cdot \cdot, T_{_{L_n}})$ gravity is that they contain up to second-order derivatives of the dynamical fields $e^a$ as in TELG, which is in contrast to the case of $f(R)$ gravity and $f(R,\mathcal{G})$ gravity \cite{Nojiri:2010wj} which contains fourth-order derivatives in the field equations. Here, $\mathcal{G}$ denotes the Gauss-Bonnet term. Useful formulae used in the derivation of the field equations are given in the appendix.

\section {Cosmological Models and Dynamical Systems Analysis}
\label{FE}

In this section we analyze the behavior of some cosmological solutions of the theory described by the action
\begin{equation}
S=\frac{1}{16 \pi G} \int_{\mathcal{M}} d^4 x e f(T_{_{GR}},T_{_{GB}}^{(1)})+ \int_{\mathcal{M}} d^4 x e \mathcal{L}_{matter} \, ,
\end{equation}
where we have fixed the space-time dimension to $D=4$. $G$ is the 4-dimensional Newton constant, $e=det(e^a_{\,\,\, \mu})= \sqrt{-g}$ and $\mathcal{L}_{matter}$ denotes a generic matter Lagrangian for matter coupled to the metric in the usual way. The matter side of the field equations is described by a perfect fluid, with density $\rho$ and pressure $P$.
So, by considering the spatially flat FLRW metric
\begin{equation}\label{metric}
ds^2=-dt^2+a(t)^2 (dx^2 +dy^2+dz^2) \,,
\end{equation}
and the simple diagonal vielbein 
\begin{equation} \label{tetrada}
e^0= dt \,, \,\,\, e^1= a(t) dx \, , \,\,\, e^2 = a(t) dy \,, \,\,\, e^3 = a(t) dz \,,
\end{equation}
we find
\begin{eqnarray}
\notag \tau_{_{GR}}=\tau_{_{L_1}} &=& -6 H^2 e^0 \wedge e^1 \wedge e^2 \wedge e^3 \\
&=& T_{_{GR}} e^0 \wedge e^1 \wedge e^2 \wedge e^3
\end{eqnarray}
from Eq. (\ref{lagrangian}) with $D=4$ and $n=1$, and 
 \begin{eqnarray}
\notag \tau_{_{GB}}^{(1)} = \tau_{_{L_2}} &=& -48 H^2 (\dot{H}+H^2)e^0 \wedge e^1 \wedge e^2 \wedge e^3 \\
&=& T_{_{GB}}^{(1)} e^0 \wedge e^1 \wedge e^2 \wedge e^3
\end{eqnarray}
from Eq. (\ref{lagrangian}) with $D=4$ and $n=2$. $H=\frac{\dot{a}(t)}{a(t)}$ is the Hubble parameter; here a dot means derivative with respect to the cosmic time $t$ .
The Friedmann equations are obtained inserting the vielbein (\ref{tetrada}) into the general field equations obtained in the previous section, and they read
\begin{eqnarray} \label{friedmann}
&& f+12 H^2 f_{T_{_{GR}}}-T_{_{GB}}^{(1)} f_{T_{_{GB}}^{(1)}}-48 H^3 \dot{f}_{T_{_{GB}}^{(1)}}= 16 \pi G \rho \, , \\
\notag && f +4(\dot{H}+3H^2)f_{T_{_{GR}}} +4H \dot{f}_{T_{_{GR}}}-T_{_{GB}}^{(1)} f_{T_{_{GB}}^{(1)}}+\frac{2}{3H} T_{_{GB}}^{(1)} \dot{f}_{T_{_{GB}}^{(1)}}-16 H^2 \ddot{f}_{T_{_{GB}}^{(1)}} =-16 \pi G P \, ,
\end{eqnarray}
where $f=f(T_{_{GR}},T_{_{GB}}^{(1)})$, $f_{T_{_{GR}}}=\frac{\partial f}{\partial T_{_{GR}}}$ and $f_{T_{_{GB}}^{(1)}}= \frac{\partial f}{\partial T_{_{GB}}^{(1)}}$.
The Friedmann equations (\ref{friedmann}) can be written in the usual form
\begin{eqnarray}\label{sol}
\notag H^2 &=& \frac{8 \pi G}{3} (\rho+\rho_{_{DE}})\,  , \\
\dot{H} &=& -\frac{ 8 \pi G}{2} (\rho+\rho_{_{DE}} +P+P_{_{DE}}) \, ,
\end{eqnarray}
where we have defined an effective dark energy density and dark energy pressure by
\begin{eqnarray} \label{syst}
\notag \rho_{_{DE}} &=& \frac{1}{16 \pi G} (6H^2-f-12H^2 f_{T_{_{GR}}} +T_{_{GB}^{(1)}} f_{T_{_{GB}}^{(1)}}+48 H^3 \dot{f}_{T_{_{GB}}^{(1)}}) \, ,\\
\notag P_{_{DE}} &=& \frac{1}{16 \pi G} (-2 (2 \dot{H}+3H^2)+f+4 (\dot{H}+3H^2)f_{T_{_{GR}}}+4H \dot{f}_{T_{_{GR}}}-T_{_{GB}}^{(1)} f_{T_{_{GB}}^{(1)}}\\
&&+\frac{2}{3H} T_{_{GB}}^{(1)} \dot{f}_{T_{_{GB}}^{(1)}}-16 H^2  \ddot{f}_{T_{_{GB}}^{(1)}}) \, .
\end{eqnarray}
The system of equations defined in (\ref{friedmann}) seems difficult to study analytically. However, we can analyze the simple limit case when the Hubble parameter is a constant
; therefore, $T_{_{GR}}=-6H^2$ and $T_{_{GB}}^{(1)} =-48 H^4$ are constants too. So, the system of Eqs. (\ref{friedmann}) is simplified to the following equations
\begin{eqnarray} \label{analityc}
\notag && f+12 H^2 f_{T_{_{GR}}}-T_{_{GB}}^{(1)} f_{T_{_{GB}}^{(1)}}= 16 \pi G \rho \, , \\
&& f +12H^2 f_{T_{_{GR}}} -T_{_{GB}}^{(1)} f_{T_{_{GB}}^{(1)}} = -16 \pi G P \, ,
\end{eqnarray}
which are consistent when $P=-\rho=constant$, i.e., the matter with the vacuum equation of state. So, given a specific $f(T_{_{GR}}, T_{_{GB}}^{(1)})$, we can solve the above equation for the constant $H$ in terms of the coupling constants of the theory; thus, we conclude that the system admits a de Sitter expansion. 
Now, we consider the specific model
$f(T_{_{GR}},T_{_{GB}}^{(1)})=T_{_{GR}}+ \alpha_1 \sqrt{T_{_{GR}}^2 + \alpha_2 T_{_{GB}}^{(1)}}$, which is the simplest model without introducing a new mass scale in the theory \cite{Kofinas:2014aka}. It is worth mentioning that we are interested in this model in order to see the differences and/or similarities of different choices of $T_{_{GB}}^{(i)}$ which, as discussed in the previous section, leads to different theories. First, we will consider the limit case $H$ constant. In this case, the analytical solution of Eqs. (\ref{analityc}) which yields a de Sitter expansion is
\begin{equation} \label{Si}
H=\left( \frac{8 \pi G \rho}{3-\alpha_1 (3+ 2\alpha_2) \left( \frac{3}{3-4 \alpha_2}\right)^{1/2}}\right)^{1/2} \,,
\end{equation}
which is real for $-3/2 < \alpha_2 <3/4$, $\alpha_1 < \sqrt{3(3-4\alpha_2)}/ (3+2\alpha_2)$ or $\alpha_2 < -3/2$, $\alpha_1 > \sqrt{3(3-4\alpha_2)}/ (3+2\alpha_2)$. For $\alpha_2=-3/2$, $H$ simplifies to $H=\left(\frac{8 \pi G \rho}{3} \right)^{1/2}$. When $\alpha_1 \rightarrow \frac{\sqrt{3(3-4 \alpha_2)}}{3+2 \alpha_2}$ the denominator of (\ref{Si}) tends to zero, so $H$ diverges unless $\rho \rightarrow 0$ in a way such that $H$ remains finite. This case corresponds to a de Sitter expansion which can happen for a pure vacuum  ($\rho = P = 0$).

Now, we realize a dynamical systems analysis to the general system of differential equations (\ref{sol}) for the particular choice of $f(T_{_{GR}}, T_{_{GB}}^{(1)})$ in order to study the cosmological  evolution of the model. In order to solve the Friedmann equations it is also necessary
to specify the equation of state parameter $\omega_{matter}= P /\rho$. In what follows, we consider a dust type of matter content $\omega_{matter}=0$. Analogously, we define an equation of state parameter for the effective dark energy by $\omega_{_{DE}}= P_{_{DE}}/ \rho_{_{DE}}$. So, by defining the variable $M=3H^2-4 \alpha_2 (\dot{H}+H^2)$ and defining also the dimensionless variables
\begin{equation}
X = \sqrt{\frac{M}{3H^2}} \, , \,\,\, Y = \frac{8 \pi G \rho}{3 H^2}\, ,
\end{equation}
the Friedmann equations (\ref{sol}) reduce to the following autonomous dynamical system
\begin{eqnarray} \label{eqs}
 \notag  X^{\prime} &=& \frac{ X (3 \alpha_1 X^2 -6(1-Y)X +\alpha_1 (3+8 \alpha_2))}{ 4 \alpha_1 \alpha_2} \,, \\
Y^{\prime} &=& \frac{3X^2-3-2 \alpha_2}{2 \alpha_2} Y\,,
\end{eqnarray}
where the prime denotes derivative with respect to $\ln a$. A useful parameter that gives information of the cosmological system is the deceleration parameter, given by
\begin{equation} \label{q}
q=-1-\frac{\dot{H}}{H^2} =-\frac{3}{4 \alpha_2} (1-X^2) \,.
\end{equation}
If $q>0$, the universe decelerates, whereas if $q<0$, then it expands accelerated. The deceleration parameter can be expressed in terms of the density parameters $\Omega_i$,  defined by $\Omega_i = \frac{8 \pi G \rho_i}{3H^2}$, as
\begin{equation}
    q=\frac{1}{2} \sum \Omega_i (1+3 \omega_i) \,,
\end{equation}
where $i$ corresponds to matter ($\omega_m=0$), radiation ($\omega_{rad}=1/3$) and dark energy, which yields
\begin{equation}
    q= \frac{1}{2} \Omega_m + \Omega_{rad}+ \frac{1}{2} \Omega_{_{DE}} (1+ 3 \omega_{_{DE}}) \, .
\end{equation}
From this equation, considering $\Omega_{rad}=0$, $\Omega_m+\Omega_{_{DE}}=1$ and Eq. (\ref{q}), we obtain
\begin{equation}
    \omega_{_{DE}}=-\frac{3X^2-3-2 \alpha_2}{6\alpha_2(Y-1)} \, ,
\end{equation}
where $\Omega_m=Y$.
The corresponding critical points of the dynamical system (\ref{eqs}) are given by
\begin{eqnarray}
\notag P_1 : X_{_C} &=& 0 \, , \,\,\,\,\, Y_{_C}=0 \\
\notag P_2 : X_{_C} &=& \frac{1}{\alpha_1}-\frac{\sqrt{1- \alpha_1^2 (1+\frac{8}{3} \alpha_2)}}{\alpha_1} \, , \,\,\,\,\, Y_{_C}=0  \\
\notag P_3 : X_{_C} &=& \frac{1}{\alpha_1}+\frac{\sqrt{1- \alpha_1^2 (1+\frac{8}{3} \alpha_2)}}{\alpha_1} \, , \,\,\,\,\, Y_{_C}=0 \\
P_4 :   X_{_C} &=& \sqrt{1+ \frac{2}{3}\alpha_2}\, , \,\,\,\,\, Y_{_C}=\frac{3 (3+2 \alpha_2)- \sqrt{3 (3+2 \alpha_2)} \alpha_1 (3+5 \alpha_2)}{3 (3+2\alpha_2)} \, .
\end{eqnarray}
The stability of the critical points can be analyzed by studying the linearized equations near the critical points, which are obtained by setting $X=X_{_C}+\delta X$ and $Y= Y_{_C} + \delta Y$ in (\ref{eqs}) and keeping only first order terms; thus, we obtain
\begin{equation}
\delta \vec{\mathbf{X}}^{\prime} = \mathbf{J} \delta \vec{\mathbf{X}} \,,
\end{equation}
where $\delta \vec{\mathbf{X}} ^{\prime} =  \left( \begin{smallmatrix} 
\delta X ^{\prime} \\ \delta Y ^{\prime}
\end{smallmatrix} \right)$, $\delta \vec{\mathbf{X}} =  \left( \begin{smallmatrix} 
\delta X \\ \delta Y
\end{smallmatrix} \right)$ and the Jacobian matrix is given by
\begin{equation}
\mathbf{J}=  \Bigg( \begin{smallmatrix}
\frac{9 X_{_C}^2}{4 \alpha_2}-\frac{3X_{_C}}{\alpha_1 \alpha_2} (1-Y_{_C})+\frac{1}{4 \alpha_2} (3+8 \alpha_2)  &  \frac{3 X_{_C}^2}{2 \alpha_1 \alpha_2}  \\ \frac{3 X_{_C} Y_{_C}}{\alpha_2} & -\frac{3-3X_{_C}^2+2 \alpha_2}{2 \alpha_2}
\end{smallmatrix} \Bigg)  \,.
\end{equation}
Now, we perform an analysis of the existence and stability of the critical points in order to study the dynamics of the system near the critical points.

\begin{itemize}

\item \underline{Critical point $P_1$}

In this case, the eigenvalues of the Jacobian matrix $\mathbf{J}$ are
\begin{equation}
 \notag \lambda_1= -1-\frac{3}{2 \alpha_2} \, , \,\,\, \lambda_2 = 3+ \frac{3}{4 \alpha_2} \, .
\end{equation}
Therefore, $P_1$ is an unstable node for $-\frac{3}{2} < \alpha_2 < -\frac{3}{8}$.

 \item \underline{Critical point $P_2$ }

The critical point $P_2$
exists for $0< \alpha_1 \leq \frac{1}{\sqrt{1+\frac{8}{3} \alpha_2}}$, $-\frac{3}{8} < \alpha_2$ or $\alpha_1 \neq 0$, $\alpha_2 = -\frac{3}{8}$ or $\alpha_1 < 0$, $\alpha_2 < - \frac{3}{8}$. The eigenvalues of the Jacobian matrix $\mathbf{J}$ are
\begin{eqnarray}
\notag \lambda_1&=& - \frac{-3+ \alpha_1^2 (3+5 \alpha_2) +\sqrt{9-3 \alpha_1^2 (3+8 \alpha_2)}}{ \alpha_2 \alpha_1^2} \, , \\
\notag \lambda_2&=& - \frac{-3+ \alpha_1^2 (3+8 \alpha_2) +\sqrt{9-3 \alpha_1^2 (3+8 \alpha_2)}}{2 \alpha_2 \alpha_1^2} \, .
\end{eqnarray}
Therefore, $P_2$ is a stable node for 
\begin{equation}
\notag    \alpha_1 < - \sqrt{\frac{3(3+2 \alpha_2)}{(3+5 \alpha_2)^2}}\, , \,\, \, - \frac{3}{5} < \alpha_2 < -\frac{3}{8}
\end{equation}    
or
\begin{equation}
 \notag    \alpha_2 < - \frac{3}{2} \, , \,\,\, \alpha_1 < 0
\end{equation}
or
\begin{equation}
\notag \alpha_2 >0 \, , \,\,\, 0 < \alpha_1 < \sqrt{\frac{3 (3 +2 \alpha_2)}{(3+ 5 \alpha_2)^2}} \, .
\end{equation}
$P_2$ is an unstable node for 
\begin{equation}
\notag    -\frac{3}{8} < \alpha_2 < 0 \, ,  \,\,\, 0 < \alpha_1 < \sqrt{\frac{1}{1+\frac{8}{3} \alpha_2}} \, .
\end{equation}

\item \underline{Critical point $P_3$}

There is a critical point $P_3$
for $0 < \alpha_1 < \frac{1}{1+\frac{8}{3} \alpha_2}$, $\alpha_2 > -\frac{3}{8}$ or $\alpha_1> 0$, $\alpha_2 \leq - \frac{3}{8}$. The eigenvalues of the Jacobian matrix $\mathbf{J}$ are
\begin{eqnarray}
\notag \lambda_1&=& \frac{3- \alpha_1^2 (3+5 \alpha_2) +\sqrt{9-3 \alpha_1^2 (3+8 \alpha_2)}}{ \alpha_2 \alpha_1^2}  \, ,\\
\notag \lambda_2&=& \frac{3- \alpha_1^2 (3+8 \alpha_2) +\sqrt{9-3 \alpha_1^2 (3+8 \alpha_2)}}{2 \alpha_2 \alpha_1^2} \, .
\end{eqnarray}
$P_3$ is a stable node for 
\begin{equation}
\notag     -\frac{3}{5} < \alpha_2 < 0 \, , \,\,\, 0 < \alpha_1 < \sqrt{\frac{3 (3+ 2\alpha_2)}{(3+5 \alpha_2)^2}} \, .
\end{equation}
$P_3$ is an unstable node for 
\begin{equation}
\notag    \alpha_2 > 0 \, , \,\,\, 0 < \alpha_1 < \sqrt{\frac{1}{1+ \frac{8}{3} \alpha_2}}  \, .
\end{equation}

\item \underline{Critical point $P_4$}

The critical point $P_4$
exists for $\alpha_2=-\frac{3}{5}$ or $-\frac{3}{2}< \alpha_2 < -\frac{3}{5}$, $\alpha_1 \geq \frac{\sqrt{3(3+2 \alpha_2)}}{3+5 \alpha_2}$ or $-\frac{3}{5} < \alpha_2$, $\alpha_1 \leq \frac{\sqrt{3 (3+2 \alpha_2)}}{3+5\alpha_2}$ . The eigenvalues of the Jacobian matrix $\mathbf{J}$ are
\begin{eqnarray}
\notag \lambda_1&=& -\frac{3}{4}- \frac{\sqrt{-\alpha_1 (72+\alpha_2 (168+71 \alpha_2))+8 \sqrt{3} (3+2\alpha_2)^{3/2}}}{4 \sqrt{\alpha_1} \alpha_2} \,,  \\
\notag \lambda_2&=& -\frac{3}{4}+ \frac{\sqrt{-\alpha_1 (72+\alpha_2 (168+71 \alpha_2))+8 \sqrt{3} (3+2\alpha_2)^{3/2}}}{4 \sqrt{\alpha_1} \alpha_2}\,.
\end{eqnarray}
$P_4$ is a stable focus for
\begin{equation}
\notag 0 > \alpha_1 > \frac{8 \sqrt{3}(3+2 \alpha_2)^{3/2}}{72+ \alpha_2(168+71 \alpha_2)} \, ,\,\,\,  \alpha_2< \frac{6}{71} (-14+3 \sqrt{6})
\end{equation}
or
\begin{equation}
\notag \alpha_1 < 0 \, , \,\,\,\,  \alpha_2 > \frac{6}{71} (-14+3 \sqrt{6}) \, .
\end{equation}

\end{itemize}

\begin{figure}[!h]
\begin{center}
\includegraphics[width=65mm]{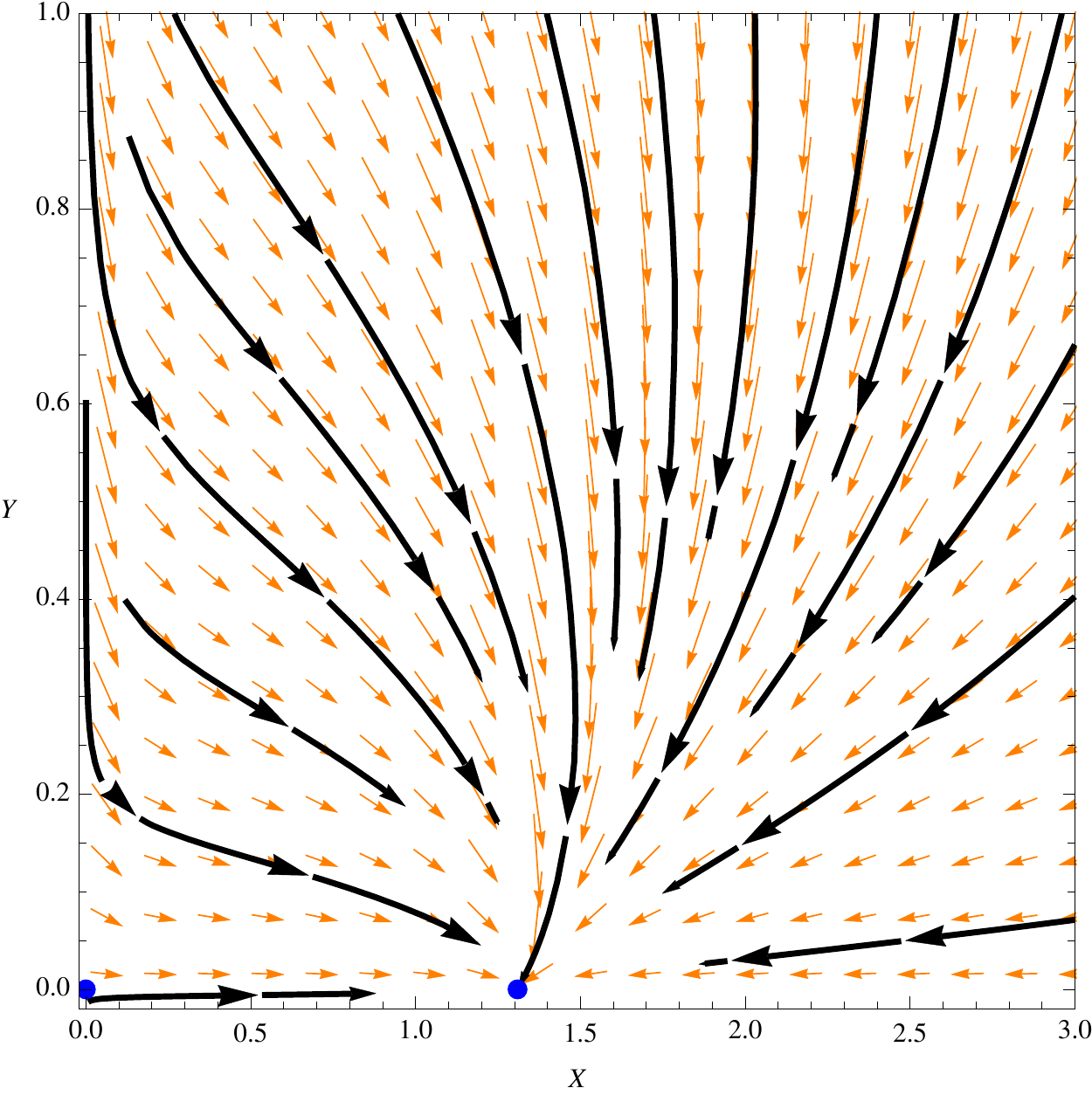}
\includegraphics[width=65mm]{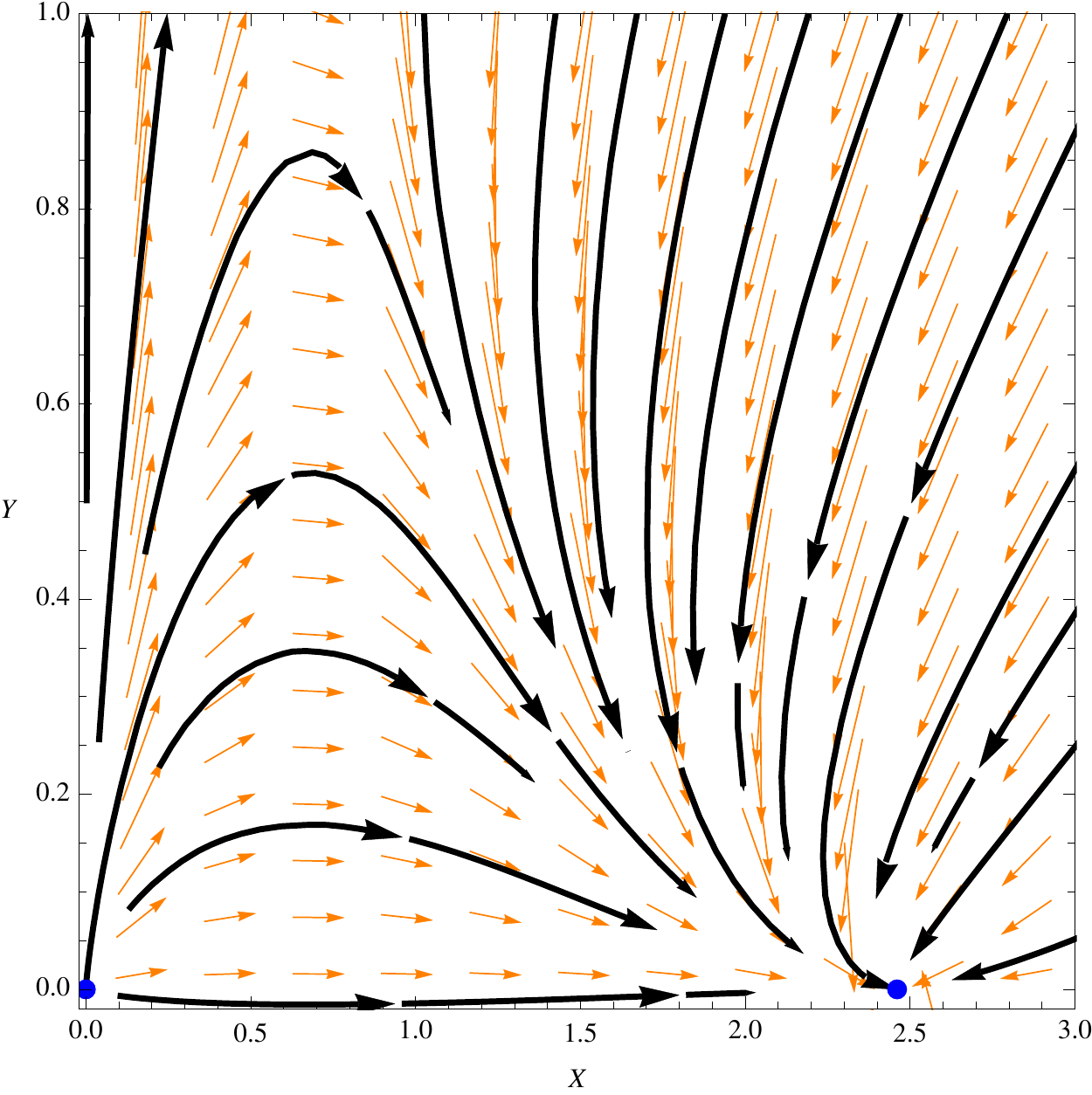}
\end{center}
\caption{Phase space for the dynamical system (\ref{eqs}). For the left panel,  $\alpha_1=-1$ and $\alpha_2=-2$, the blue points corresponds to the critical points $P_1=(0,0)$ and $P_2=(1.31,0)$, the deceleration parameters are $q(P_1)=0.38$ and $q(P_2)=-0.27$, and $\omega_{_{DE}}(P_1)=-0.08$, $\omega_{_{DE}}(P_2)=-0.51$. For the right panel, $\alpha_1=1$ and $\alpha_2=-0.8$, the blue points corresponds to the critical points $P_1=(0,0)$ and $P_3=(2.46,0)$, the deceleration parameters are $q(P_1)=0.94$ and $q(P_3)=-4.74$,  and $\omega_{_{DE}}(P_1)=0.29$, $\omega_{_{DE}}(P_3)=-3.49$.}
\label{Fig}
\end{figure}

\begin{figure}[!h]
\begin{center}
\includegraphics[width=65mm]{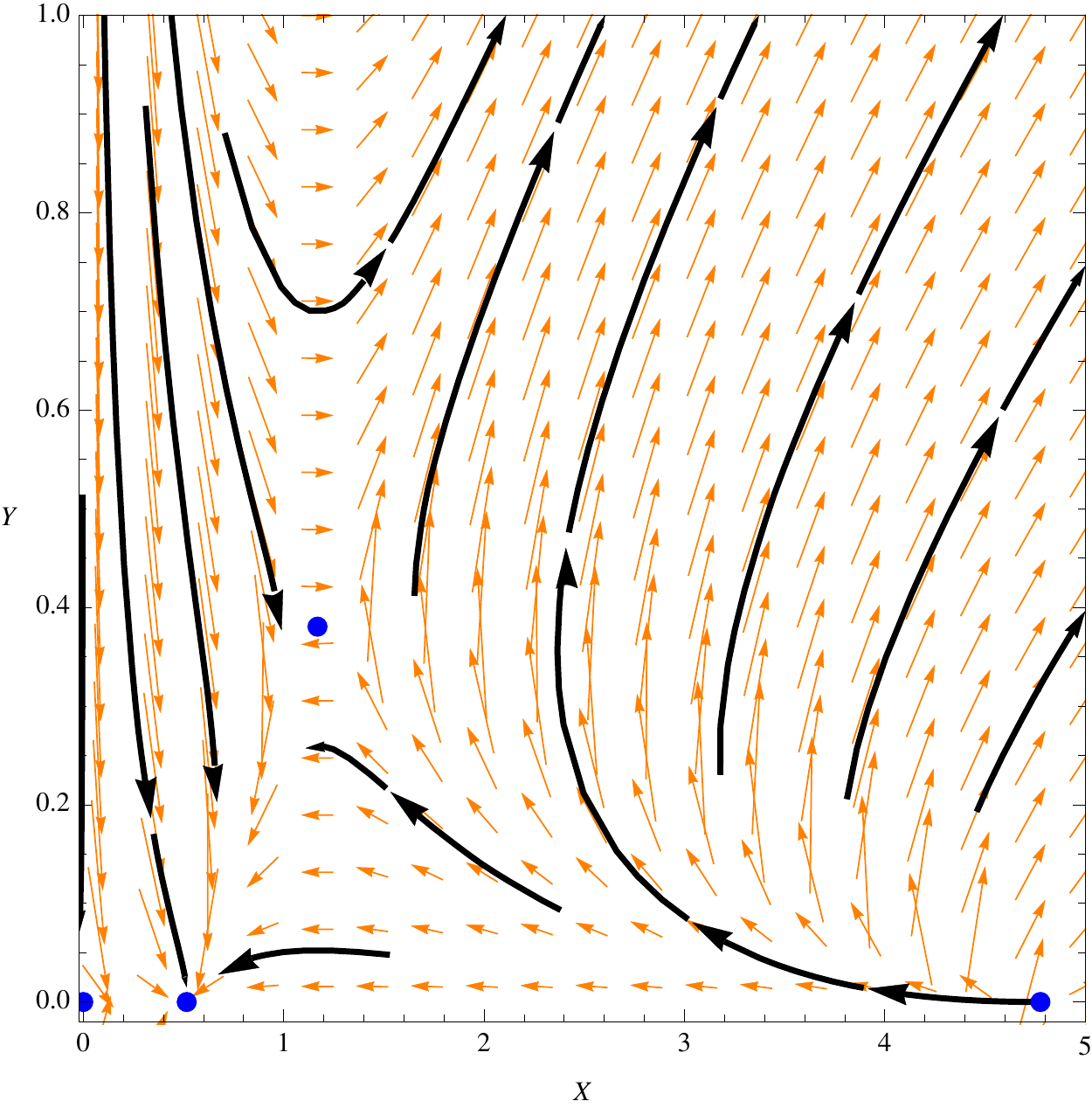}
\includegraphics[width=65mm]{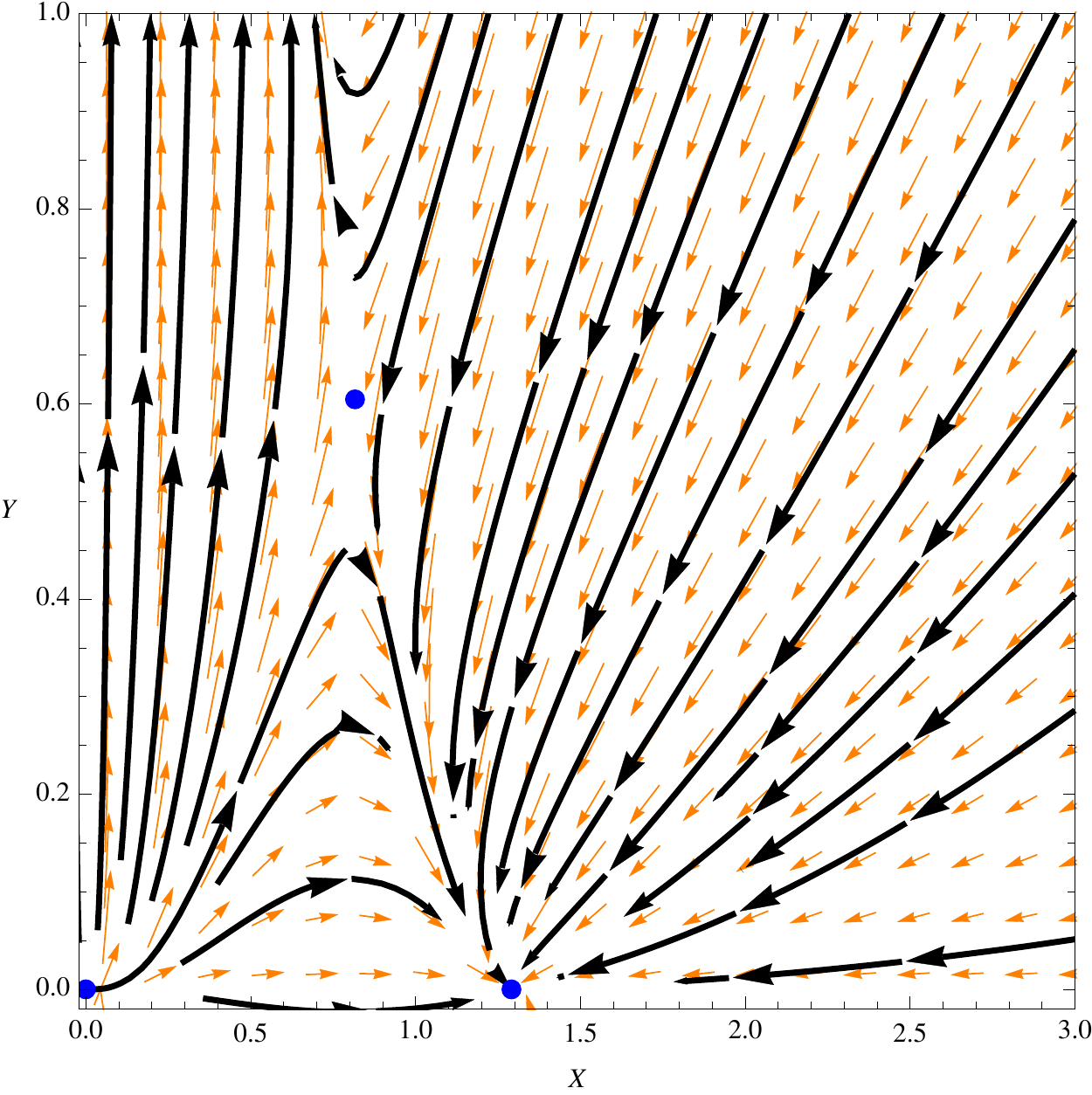}
\end{center}
\caption{Phase space for the dynamical system (\ref{eqs}). For the left panel,  $\alpha_1= \frac{\sqrt{3(3-4 \alpha_2)}}{3+2 \alpha_2}$ and $\alpha_2=0.55$, the blue points corresponds to the critical points $P_1=(0,0)$, $P_2=(0.52,0)$, $P_3=(4.78,0)$ and $P_4=(1.17,0.38)$, the deceleration parameters are $q(P_1)=-1.36$, $q(P_2)=-1$, $q(P_3)=29.75$ and $q(P_4)=0.5$, and $\omega_{_{DE}}(P_1)=-1.24$, $\omega_{_{DE}}(P_2)=-1$,$\omega_{_{DE}}(P_3)=19.50$, $\omega_{_{DE}}(P_4)=0$. For the right panel, $\alpha_1= \frac{\sqrt{3(3-4 \alpha_2)}}{3+2 \alpha_2}$ and $\alpha_2=-0.5$, the blue points corresponds to the critical points $P_1=(0,0)$, $P_3=(1.29,0)$ and $P_4=(0.82,0.60)$, the deceleration parameters are $q(P_1)=1.5$, $q(P_3)=-1$ and $q(P_4)=0.5$,  and $\omega_{_{DE}}(P_1)=0.67$, $\omega_{_{DE}}(P_3)=-1$, $\omega_{_{DE}}(P_4)=0$.}
\label{Fig1}
\end{figure}

Now, in Fig. (\ref{Fig}) we plot the phase space portrait for different choices of the parameters $\alpha_1$ and $\alpha_2$. In the left panel there are two critical points $P_1=(0,0)$ and $P_2=(1.31,0)$, and the deceleration parameters are $q(P_1)=0.38$ and $q(P_2)=-0.27$. We have also computed the equation of state parameter for dark energy, which yields $\omega_{_{DE}}(P_1)=-0.08$, $\omega_{_{DE}}(P_2)=-0.51$. For the right panel, the critical points are $P_1=(0,0)$ and $P_3=(2.46,0)$, the deceleration parameters are $q(P_1)=0.94$ and $q(P_3)=-4.74$,  and $\omega_{_{DE}}(P_1)=0.29$, $\omega_{_{DE}}(P_3)=-3.49$. In Fig. (\ref{Fig1}) we plot the phase space portrait for $\alpha_1=\frac{\sqrt{3(3-4 \alpha_2)}}{3+2 \alpha_2}$ and different choices of $\alpha_2$. In the left panel there are four critical points $P_1=(0,0)$, $P_2=(0.52,0)$, $P_3=(4.78,0)$ and $P_4=(1.17,0.38)$, and the deceleration parameters are $q(P_1)=-1.36$, $q(P_2)=-1$, $q(P_3)=29.75$ and $q(P_4)=0.5$, also we have computed the equation of state parameter for dark energy, which yields  $\omega_{_{DE}}(P_1)=-1.24$, $\omega_{_{DE}}(P_2)=-1$, $\omega_{_{DE}}(P_3)=19.50$, $\omega_{_{DE}}(P_4)=0$. For the right panel, the critical points are $P_1=(0,0)$, $P_3=(1.29,0)$ and $P_4=(0.82,0.60)$, the deceleration parameters are $q(P_1)=1.5$, $q(P_3)=-1$ and $q(P_4)=0.5$,  and $\omega_{_{DE}}(P_1)=0.67$, $\omega_{_{DE}}(P_3)=-1$, $\omega_{_{DE}}(P_4)=0$. 
We observe that depending on the values of the parameters, some critical points can attract the universe at late times, and the corresponding values of $q$ and $\omega_{_{DE}}$ can lie in the quintessence regime, cosmological constant regime or phantom regime.
Interestingly, we note that the
behavior of the system, critical points and stability can be identified with the behavior obtained in \cite{Kofinas:2014aka} for the theory $f(T_{_{GR}},T_{_{GB}}^{(2)})$ by a re-scaling of the dimensionless coupling constant $\alpha_2 \rightarrow -\frac{\alpha_2}{2}$. This is because in that case our equations are similar to that of \cite{Kofinas:2014aka}; therefore, we conclude that the theory $f(T_{_{GR}},T_{_{GB}}^{(1)})$ can produce similar results to the theory $f(T_{_{GR}},T_{_{GB}}^{(2)})$, but in different regions of the parameter space, as we have shown explicitly above. We suspect a similar behavior can occur for the other Lagrangians $T_{GB}^{(3)}$ and $T_{GB}^{(4)}$ and the flat FLRW metric. Certainly, the same can also happen by selecting other choices of the function $f$.

\section{Conclusions}
\label{C}

In this work we studied TELG \cite{Gonzalez:2015sha} and its natural extension
by considering an arbitrary function $f(T_{_{L_1}}, T_{_{L_2}}, \cdot \cdot \cdot , T_{_{L_n}})$ of the torsion invariants $T_{_{L_n}}$ of degree $2n$ in the torsion tensor that appear in the action of TELG, which are also topological invariants in $2n$ dimensions. 
We derived the field equations of this gravitational theory and found that they contain up to second-order derivatives of the dynamical fields $e^a$ as happens in TELG, which is in contrast for example to the case of $f(R)$ gravity and $f(R,\mathcal{G})$ gravity which contains fourth order derivatives in the field equations; here $\mathcal{G}$ denotes the Gauss-Bonnet term. As application, we studied some cosmological implications of the theory by considering the special case of a natural extension of teleparallel equivalent of Gauss-Bonnet gravity in four space-time dimensions, where the function $f(T_{_{GR}},T_{_{GB}}^{(1)})$ depends only on the invariants $T_{_{GR}}$ and $T_{_{GB}}^{(1)}$. We considered the spatially flat FLRW metric and found the Friedmann equations, then we chose a specific model given by $f(T_{_{GR}},T_{_{GB}}^{(1)})= T_{_{GR}}+ \alpha_1 \sqrt{T_{_{GR}}^2+ \alpha_2 T_{_{GB}}^{(1)}}$. Then, we performed a dynamical systems analysis to the cosmological model, and we found that depending on the parameters $\alpha_1$ and $\alpha_2$, some critical points can attract the universe at late times, and the corresponding values of $q$ and $\omega_{_{DE}}$ can lie in the quintessence regime, cosmological constant regime or phantom regime. The theory analyzed here is different from the theory 
considered in \cite{Kofinas:2014owa}. In fact, by performing integration by parts to $T_{_{GB}}^{(1)}$, it is possible to
obtain four Lagrangians $T_{_{GB}}^{(i)}$ ($i=1,2,3,4$) differing from each other by boundary terms \cite{Gonzalez:2015sha}, thus yielding field equations equivalent to the Gauss-Bonnet gravity. The Lagrangian analyzed here $T_{_{GB}}^{(1)}$ is the simplest of them, while the one considered in \cite{Kofinas:2014owa} was $T_{_{GB}}^{(2)}$ according to the classification given in \cite{Gonzalez:2015sha}; however, by considering an arbitrary function of the torsion invariants, the modified Lagrangians $f(T_{_{GR}},T_{_{GB}}^{(i)})$ define different gravitational theories due to the equations of motion of the theories not being equivalent, 
except in the special case when $f(T_{_{GR}},T_{_{GB}}^{(i)})=T_{_{GR}}+\alpha T_{_{GB}}^{(i)}$, where the theories are also equivalent to the Einstein-Gauss-Bonnet gravity. 
The model analyzed was studied in \cite{Kofinas:2014aka}, but in the theory $f(T_{_{GR}},T_{_{GB}}^{(2)})$. We chose the same model in order to see the similarities and/or differences with respect to the theory $f(T_{_{GR}},T_{_{GB}}^{(1)})$, which is a special case of $f(T_{_{L_1}}, T_{_{L_2}}, \cdot \cdot \cdot, T_{_{L_n}})$. Interestingly, we found that despite the theories being different, our solutions can be identified with the solutions obtained in \cite{Kofinas:2014aka} by a re-scaling of the dimensionless coupling constant $\alpha_2 \rightarrow -\frac{\alpha_2}{2}$, and we suspect a similar behavior can occur for the other Lagrangians. Therefore, at least for the case of the flat FLRW metric analyzed here, the theories can produce similar results but in different regions of the space of parameters. We hope to study other cosmological models by incorporating into $f$ torsion invariants of higher degree in the torsion tensor.

\section*{Acknowledgments}

We would like to thank the anonymous referee for valuable comments which help us to improve the quality of our paper. We would also like to thank to Mar\'ia Jos\'e Guzm\'an for helpful suggestions and comments. This work was partially funded by the Comisi\'{o}n
Nacional de Ciencias y Tecnolog\'{i}a through FONDECYT Grants
11140674 (P.A.G. and S.R.) and by the Direcci\'{o}n de Investigaci\'{o}n y Desarrollo de la Universidad de La Serena (S.R. and Y.V.). P. A. G. acknowledges the hospitality of the Universidad de La Serena where part of this work was undertaken.

\section{Appendix}
In this appendix we present some formulae used in the derivation of the field equations, which can be found in \cite{Kofinas:2014owa}. However, we shall consider from the beginning that $\omega^{ab}=0$.
We denote with a bold letter the vector field $\mathbf{e}_{a}$ which forms an orthonormal basis for the tangent space at each point of the manifold, while $e_{a}= \eta_{ab} e^{b}$ are 1-forms.

Through consecutive application of the interior product $i_{\mathbf{e}_a}$ to the relation $T^{a}=K^{a}_{\,\,\, b} \wedge e^{b}$, the contortion 1-form can be expressed in terms of the torsion 2-form as
\begin{equation}
\notag    K_{ab}=\frac{1}{2} \left( i_{\mathbf{e}_a} T_b -i_{\mathbf{e}_b} T_a -(i_{\mathbf{e}_a} i_{\mathbf{e}_b} T_c) \wedge e^{c} \right)\, ;
\end{equation}
thus, the variation of the contortion with respect to the vielbein yields
\begin{equation}
\notag \delta K_{ab}= \frac{1}{2} \left( \delta (  i_{\mathbf{e}_a} T_b ) -\delta ( i_{\mathbf{e}_b} T_a) -\delta( (i_{\mathbf{e}_a} i_{\mathbf{e}_b} T_c) ) \wedge e^{c} -(i_{\mathbf{e}_a} i_{\mathbf{e}_b} T_c) \wedge \delta e^{c} \right) \,.
\end{equation}
Using the relations \cite{Kofinas:2014owa}
\begin{eqnarray}
\notag \delta (i_{\mathbf{e}_a}T^{b}) &=& \mathcal{L}_{\mathbf{e}_a} \delta e^b+ \mathcal{L}_{\mathbf{e}_c}(e^b(\delta \mathbf{e}_a)) \wedge e^c+ C^{b}_{\,\,\, cd} e^{d} (\delta \mathbf{e}_a) \wedge e^c \, ,\\
\notag \delta (i_{\mathbf{e}_a} i_{\mathbf{e}_b} T^{c})&=& \mathcal{L}_{\mathbf{e}_a}(e^{c} (\delta \mathbf{e}_b)) - \mathcal{L}_{\mathbf{e}_b}(e^{c} (\delta \mathbf{e}_a))+ C^{c}_{\,\,\, ad} e^{d} (\delta \mathbf{e}_{b})-C^{c}_{\,\,\, bd} e^{d} (\delta \mathbf{e}_{a})-C^{d}_{\,\,\, ab} e^{c} (\delta \mathbf{e}_{d}) \, ,
\end{eqnarray}
the following expression can be obtained
\begin{eqnarray}
\notag \delta K_{ab} &=& \frac{1}{2} \Big(  \mathcal{L}_{\mathbf{e}_a} \delta e_b -\mathcal{L}_{\mathbf{e}_b} \delta e_a  +\mathcal{L}_{\mathbf{e}_c}(i_{\mathbf{e}_b}(\delta e_a)) \wedge e^c -\mathcal{L}_{\mathbf{e}_c}(i_{\mathbf{e}_a}(\delta e_b)) \wedge e^c + \mathcal{L}_{\mathbf{e}_a}(i_{\mathbf{e}_b}(\delta e_c)) \wedge e^c \\
\notag && -\mathcal{L}_{\mathbf{e}_b}(i_{\mathbf{e}_a}(\delta e_c)) \wedge e^c - C^{d}_{\,\,\, ab}(i_{\mathbf{e}_d} \delta e_{c})\wedge e^{c}+2 C_{(ac)d}(i_{\mathbf{e}_b} \delta e^{d})\wedge e^{c} - 2 C_{(bc)d}(i_{\mathbf{e}_a} \delta e^{d}) \wedge e^{c} +T_{cab} \delta e^{c}\Big) \, .
\end{eqnarray}

\vskip 7cm

\end{document}